\documentclass{iaus}
\usepackage{graphicx}

\def\eps@scaling{1.14}
\newcommand\plottwo[2]{{%
\typeout{Plottwo included the files #1 #2}
\centering
\leavevmode
\columnwidth=.5\columnwidth
\includegraphics[width={\eps@scaling\columnwidth}]{#1}%
\hfil
\hspace{-1.5cm}
\includegraphics[width={\eps@scaling\columnwidth}]{#2}%
}}%

\title[Cosmic evolution of stellar disk truncations] 
{Evolution of stellar disk truncations since z=1}

\author[Ignacio Trujillo et al.]   
{Ignacio Trujillo$^1$, Ruyman Azzollini$^1$, Judit Bakos$^1$, John Beckman$^1$
 \and Michael Pohlen$^2$}

\affiliation{$^1$Instituto de Astrof\'isica de Canarias,\\ 
C/V\'ia L\'actea s/n, 38205 La Laguna, S/C de Tenerife, Spain\\ 
email: {\tt trujillo@iac.es,ruyman@iac.es,jbakos@iac.es,jeb@iac.es} \\
[\affilskip]
$^2$Cardiff University, School of Physics \& Astronomy,\\
 Cardiff, CF24 3AA, Wales, UK
 \\email: {\tt Michael.Pohlen@astro.cf.ac.uk}}

\pubyear{2008}
\volume{xxx}  
\pagerange{119--126}
\jname{Title of your IAU Symposium}
\editors{A.C. Editor, B.D. Editor \& C.E. Editor, eds.}
\begin{document}

\maketitle

\begin{abstract} 

We present our recent results on the cosmic evolution of the outskirst of disk
galaxies. In particular we focus on disk--like galaxies with stellar disk
truncations. Using UDF, GOODS and SDSS data we show how the position of the
break (i.e. a direct estimator of the size of the stellar disk) evolves with
time since z$\sim$1. Our findings agree with an evolution on the radial position
of the break by a factor of 1.3$\pm$0.1 in the last 8 Gyr for galaxies with
similar stellar masses. We also present radial color gradients and how they
evolve with time. At all redshift we find  a radial inside-out bluing reaching a
minimum  at the position of the break radius, this minimum is followed by a
reddening outwards. Our results constraint several galaxy disk formation models
and favour a scenario where stars are formed inside the break radius and are
relocated in the outskirts of galaxies through secular processes.

\end{abstract}

\firstsection 
\section{Introduction}

Early studies of the disks of spiral galaxies (\cite[Patterson
1940]{Patterson40}, \cite[de Vaucouleurs 1959]{deVaucouleurs59}, \cite[Freeman
1970]{Freeman70}) showed that this component generally follows an exponential
radial surface-brightness profile, with a certain scale length, usually taken as
the characteristic size of the disk. \cite[Freeman (1970)]{Freeman70} pointed
out, though, that not all disks follow this simple exponential law. In fact, a
repeatedly reported feature of disks for a representative fraction of the spiral
galaxies is that of a truncation of the stellar population at large radii,
typically 2-4 exponential scale lengths (see e.g. the review
by \cite[Pohlen et al. 2004]{Pohlen04}).

 Several possible break-forming mechanisms have been investigated to explain the truncations. There
have been ideas based on maximum angular momentum distribution: \cite[van der Kruit (1987)]{Kruit87}
proposed that angular momentum conservation in a collapsing, uniformly rotating cloud naturally gives
rise to disk breaks at roughly 4.5 scale radii. \cite[van den Bosch (2001)]{Bosch01} suggested that the
breaks are due to angular momentum cut-offs of the cooled gas. On the other hand, breaks have also been
attributed to a threshold for star formation (SF), due to changes in the gas density \cite[Kennicutt
(1989)]{Kennicutt89}, or to an absence of equilibrium in the cool Interstellar Medium phase
(\cite[Elmegreen \& Parravano 1994]{ElmegreenParravano94}, \cite[Schaye 2004]{Schaye04}). Magnetic
fields have been also considered (Battaner et al. 2002) as responsible of the truncations. More recent
models using collisionless N-body simulations, such as that by \cite[Debattista et al.
(2006)]{Debattista06}, demonstrated that the redistribution of angular momentum by spirals during bar
formation also produces realistic breaks. In a further elaboration of this idea, \cite[Ro\v{s}kar et
al. (2008)]{Roskar08} have performed high resolution simulations of the formation of a galaxy embedded
in a dark matter halo. In these models, breaks are the result of the interplay between a radial star
formation cut-off and redistribution of stellar mass by secular processes. A natural prediction of
these models is that the stellar populations present an age minimum in the break position. This
prediction could be probed by exploring the color profiles of the galaxies.

Furthermore, addressing the question of how the radial truncation evolves with z
is strongly linked to our understanding of how the galactic disks grow and where
star formation takes place. \cite[P\'erez (2004)]{Perez04} showed that it is
possible to detect stellar truncations even out to z$\sim$1. Using the radial
position of the truncation  as a direct estimator of the size of the stellar
disk, \cite[Trujillo \& Pohlen (2005)]{TP05} inferred a moderate ($\sim $25\%)
inside-out growth of disk galaxies since z$\sim$1. An important point, however
was missing in the previous analyses: the evolution with redshift of the radial
position of the break at a given stellar mass. The stellar mass is a much better
parameter to explore the growth of galaxies, since the luminosity evolution of
the stellar populations can mimic a size evolution (\cite[Trujillo et al.
2004]{Trujillo04}, \cite[Trujillo et al. 2006]{Trujillo06}). We present in this
contribution a quick summary of our recent findings on the stellar disk
truncation origin and its evolution with redshift. The results presented here
are based on the following publications: \cite[Azzollini et al. (2008a)]{Azzollini08a},
\cite[Azzollini et al. (2008b)]{Azzollini08b} and \cite[Bakos et al. (2008)]{Bakos08}.
Throughout, we assume a flat $\Lambda$-dominated cosmology ($\Omega_{M}$ = 0.30,
$\Omega_{\Lambda}$=0.70, and $H_{0}$=70 km $s^{-1}$ $Mpc^{-1}$).

\section{Color profiles in Local Galaxies}

\begin{figure}
\includegraphics[width=1\textwidth]{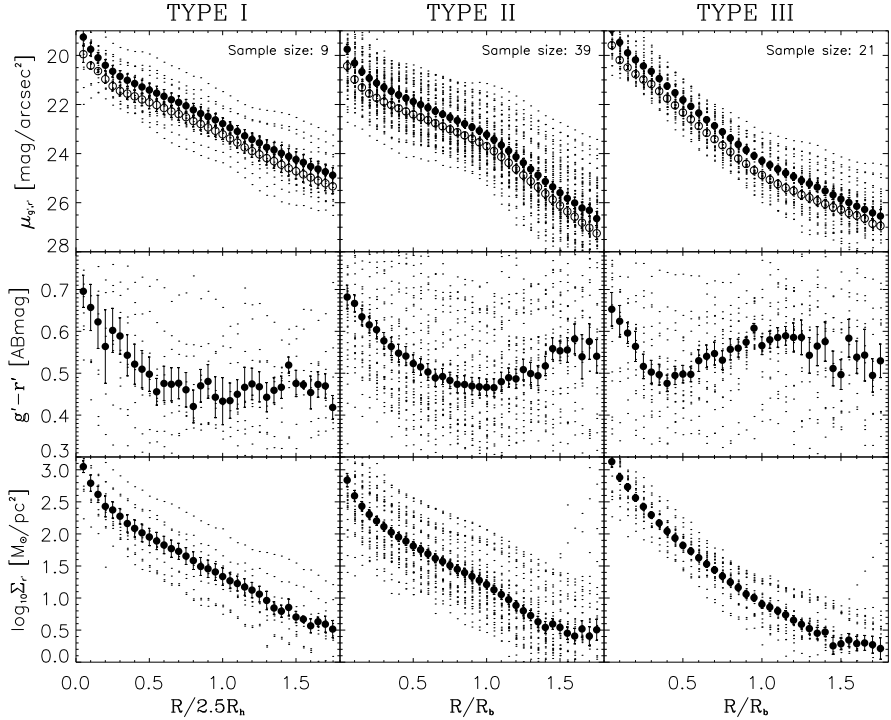}

\caption{\textit{Upper row}: Averaged, scaled radial surface brightness profiles of 9 Type~I
  (pure exponential profiles), 39 Type~II (truncated galaxies) and 21 Type~III (antitruncated)
  galaxies. The filled circles correspond to the r' band mean surface brightness, the open
  circles to the mean g' band data (\cite[Pohlen \& Trujillo 2006]{PT06}). The small dots are
  the individual galaxy profiles in both bands. The surface brightness is corrected for Galactic
  extinction. | \textit{Middle row}: (g' - r') color gradients. The averaged profile of Type~I
  reaches an asymptotic color value of $\sim 0.46 \ {\rm mag}$ being rather constant outwards.
  Type~II profiles have a minimum color of $0.47 \pm 0.02 \ {\rm mag}$ at the break position.
  The mean color profile of Type~III has a redder value of about $0.57 \pm 0.02 \ {\rm mag}$ at
  the break. | \textit{Bottom row}: r' band surface mass density profiles obtained using the
  color to M/L conversion of Bell et al. (2003). Note how the significance of the break almost
  disappears for the Type~II (truncated galaxies) case.}

\label{colorbak}
\end{figure}

In order to contrain the outer disk formation models, in Bakos et al. (2008), we have explored
radial color and stellar surface mass density profiles for a sample of 85 late-type spiral
galaxies with available deep (down to $\sim$27 mag/arcsec$^2$) SDSS g' and r' band surface
brightness profiles \cite[(Pohlen \& Trujillo 2006)]{PT06}. About $90 \% $ of the light profiles
have been classified as broken exponentials, either exhibiting truncations (Type~II galaxies) or
antitruncations (Type~III galaxies). Their associated color profiles show a significantly
different behavior. For the truncated galaxies a radial inside-out bluing reaches a minimum of
$(g'-r') = 0.47 \pm 0.02$ mag at the position of the break radius, this minimum is followed by a
reddening outwards (see middle row in Fig. \ref{colorbak}). The antitruncated galaxies reveal a
different behavior. Their break in the light profile resides in a plateau region of the color
profile at about $(g'-r') = 0.57 \pm 0.02$.

Using the $(g'-r')$ color \cite[(Bell et al. 2003)]{Bell03} to calculate the stellar surface mass
density profiles reveals a surprising result. The breaks, well established in the light profiles of the
Type~II galaxies, are almost gone, and the mass profiles resemble now those of the pure exponential
Type~I galaxies (see bottom row in Fig. \ref{colorbak}). This result suggests that the origin of the
break in Type~II galaxies is more likely due to a radial change in the ingredients of the stellar
population than being associated to an actual drop in the distribution of mass. The antitruncated
galaxies on the other hand preserve to some extent their shape in the stellar mass density profiles. 

There are other structural parameters that can be computed to contrain the different formation
scenarios. Among these we have estimated the stellar surface mass density at the break for
truncated (Type~II) galaxies (13.6$\pm$1.6 M$_{\odot}{pc}^{-2}$) and the same parameter for the
antitruncated (Type~III) galaxies (9.9$\pm$1.3 M$_{\odot}{pc}^{-2}$). Finally,  we have measured
that $\sim$15\% of the total stellar mass in case of truncated galaxies and $\sim$9\% in case
of antitruncated galaxies are to be found beyond the measured break radii in the light profiles.

\section{Stellar disk truncation evolution}

\begin{figure}
\plottwo{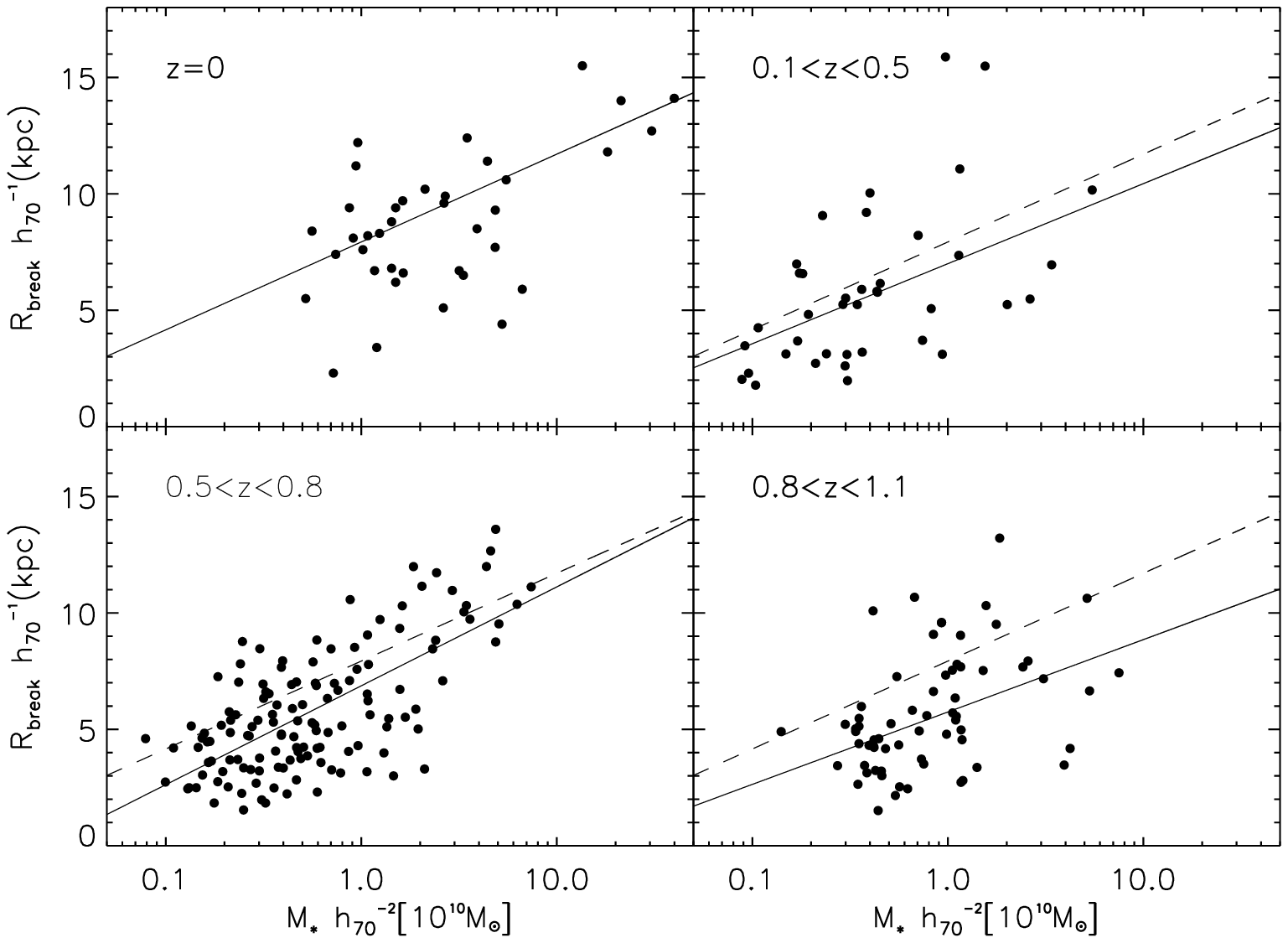}{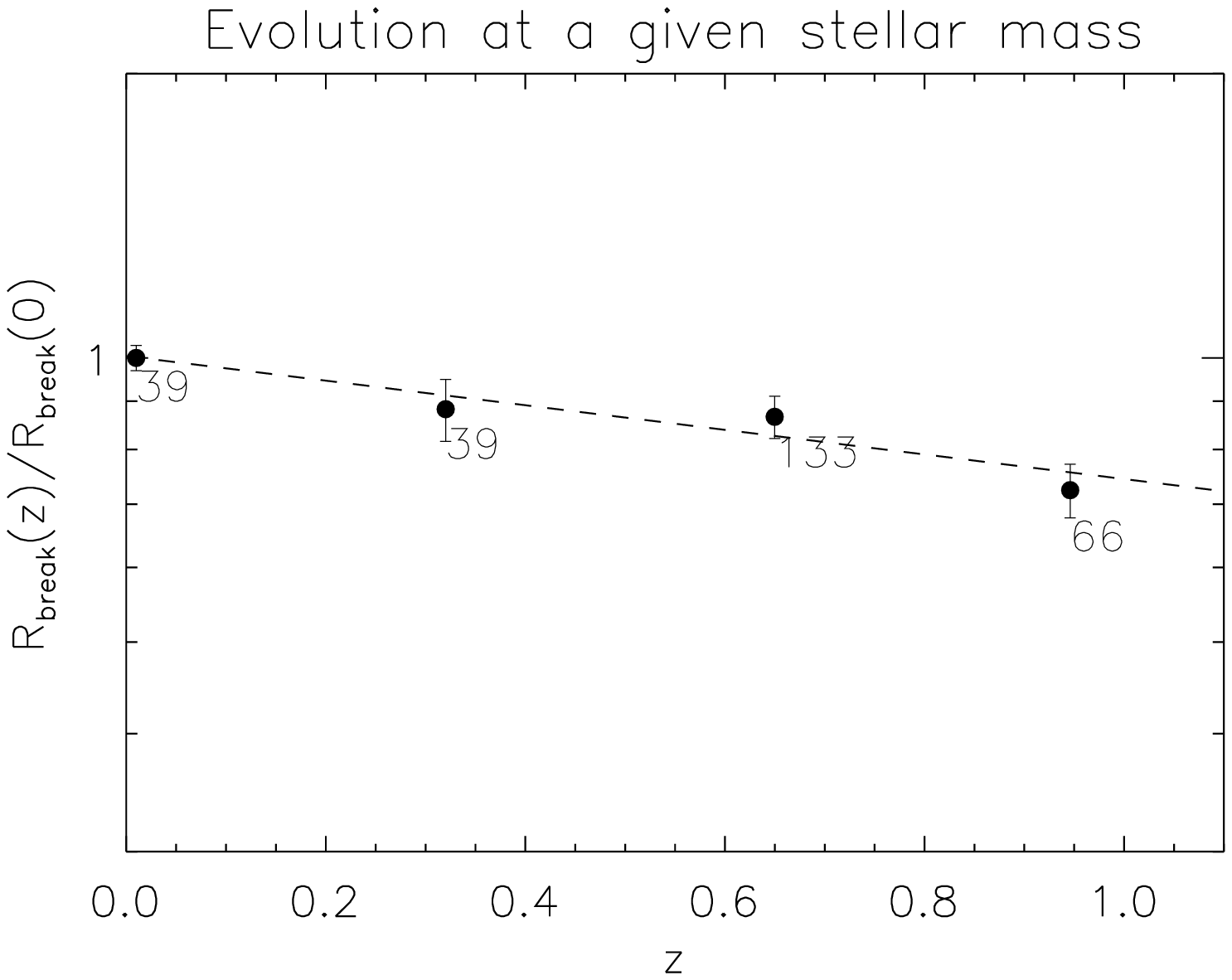}

\caption{Left: Break Radius of ``truncated'' galaxies as a function of stellar mass, for 4
ranges of redshift. Local data are from \cite[Pohlen \& Trujillo (2006)]{PT06} ($g'$-band
results). Right: Size evolution at a given stellar mass of the break radius as a function of
redshift. We have found  a growth of a factor 1.3$\pm$0.1 between z=1 and z=0. The numbers
acompanying each point in the right panel give the population of objects which they represent.}

\label{figTrmass}
\end{figure}

In Azzollini et al. (2008a), we have conducted the largest systematic search so far for stellar
disk truncations in disk-like galaxies at intermediate redshift ($z$$<$1.1), using the Great
Observatories Origins Deep Survey South (GOODS-S) data from the \emph{Hubble Space Telescope} -
ACS. Focusing on Type II galaxies (i.e. downbending profiles) we explore whether the position of
the break in the rest-frame $B$-band radial surface brightness profile (a direct estimator of
the extent of the disk where most of the massive star formation is taking place), evolves with
time. The number of galaxies under analysis (238 of a total of 505) is an order of magnitude
larger than in previous studies. For the first time, we probe the evolution of the break radius
for a given stellar mass (a parameter well suited to address evolutionary studies). Our results
suggest that, for a given stellar mass, the radial position of the break has increased with
cosmic time by a factor 1.3$\pm$0.1 between $z$$\sim$1 and $z$$\sim$0 (see Fig.
\ref{figTrmass}). This is in agreement with a moderate inside-out growth of the disk galaxies in
the last $\sim$ 8 Gyr. In the same period of time, the surface brightness level in the
rest-frame $B$-band at which the break takes place has increased by 3.3$\pm$0.2 mag/arcsec$^2$ (a
decrease in brightness by a factor of 20.9$\pm$4.2).

\begin{figure}
\includegraphics[width=0.8\textwidth]{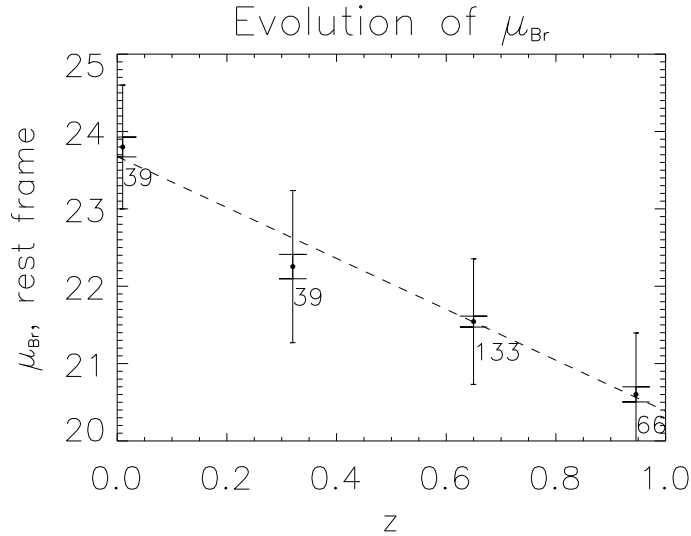}

\caption{Evolution of the surface brightness at the break for Type II galaxies with redshift. We
show the median surface brightness at the break for the distribution of our galaxies. The larger
error bars represent the standard deviation of the distributions, while the shorter ones give
the error in the median values. The numbers acompanying each point in the right panel give the
population of objects which they represent.}

\label{figmu}
\end{figure}

In Azzollini et al. (2008a) we also find that at a given stellar mass, the scale lengths of the
disk in the part inner to the ``break'' were on average somewhat larger in the past, and have
remained more or less constant until recently. This phenomenon could be related to the spatial
distribution of star formation, which seems to be rather spread over the disks in the images. So
disk galaxies had profiles with a flatter brightness distribution in the inner part of the disk,
which has grown in extension, while becoming fainter and ``steeper'' over time. This is
consistent with at least some versions of the inside-out formation scenario for disks.

\section{Color profiles in intermediate redshift galaxies}

In addition to the evolution on the position of the break in spiral galaxies is important to
explore how the color of the surface brightness profiles has evolved with time. This kind of
analysis sheds light on when stars formed in different parts of the disk of galaxies, thus
giving hints on the stellar mass buildup process.

In Azzollini et al. (2008b) we present deep color profiles for a sample of 415 disk galaxies
within the redshift range 0.1$\leq$z$\leq$1.1 , and contained in HST ACS imaging of the
GOODS-South field. For each galaxy, passband combinations are chosen to obtain, at each
redshift, the best possible approximation to the rest-frame $u-g$ color. We find that objects
which show a truncation in their stellar disk (type II objects) usually show a minimum in their
color profile at the break, or very near to it, with a maximum to minimum amplitude in color of
$\leq$0.2 mag/arcsec$^2$, a feature which is persistent through the explored range of redshifts
(i.e., in the last $\sim$8 Gyr and that it is also found in our local sample for comparison
(Bakos et al. 2008)). This color structure is in qualitative agreement with recent model
expectations where the break of the surface brightness profiles is the result of the interplay
between a radial star formation cutoff and a redistribution of stellar mass by secular processes
(\cite [Ro\v{s}kar et al. 2008]{Roskar08}). Our results fit qualitatively their prediction that
the youngest stellar population should be found at the break radius, and older (redder) stars
must be located beyond that radius. It is not easy to understand how ''angular momentum'' or
''star formation threshold''/''ISM phases'' models alone could explain our results. Thus they
pose a difficult challenge for these models. However, it will also be necessary to check whether
the Ro\v{s}kar et al. (2008) models (as well as other available models in the literature like
those of \cite[Bournaud et al. 2007]{Bour07} and \cite[Foyle et al. 2008]{Foyle08}) are able to reproduce quantitatively the
results shown here.

Combining the results found in Azzollini et al. (2008b) and Bakos et al. (2008)  one is tempted
to claim that both the existence of the break in Type~II galaxies, as well as the shape of their
color profiles, are long lived features in the galaxy evolution. Because it would be hard to
imagine how the above features could be continuously destroyed and re-created maintaining the
same properties over the last $\sim$8 Gyr.

\begin{figure}
\includegraphics[width=1\textwidth]{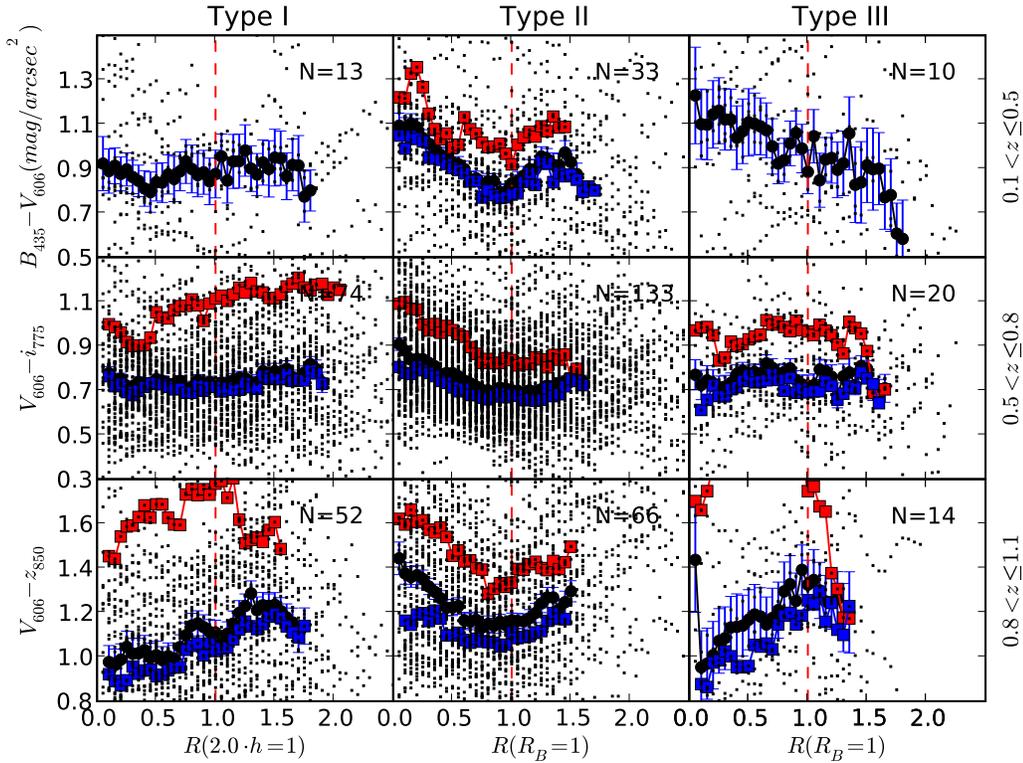}

\caption{Color profiles of the 415 galaxies under study in Azzollini et al. (2008b). The sample
 is divided in subsamples according to surface brightness profile type (I or pure exponential
 profiles, II or truncated galaxies, III or antitruncated, in columns, from left to right) and
 redshift range (low, mid, or high, in rows, from top to bottom). The colors
 ($B_{435}-V_{606}$,$V_{606}-i_{775}$,$V_{606}-z_{850}$) are chosen as the best proxies to the
 rest-frame $u-g$  color in each redshift bin. The radii are scaled to the scale radius, $R_s$,
 whose definition depends on profile type: $R_s=2h$ for type I, where h is the scale length of
 the disk, and it is equal to the break radius, $R_s=R_B$, for types II and III. Small points
 are individual color profiles. Large black dots are the median color profiles for each
 subsample, and the error bars give the error in those estimations. The red squares give the
 median color profile for objects with stellar mass M$_{\star}$$>$10$^{10}$M$_{\odot}$, while the
 blue squares give the same for objects with M$_{\star}$$\leq$10$^{10}$M$_{\odot}$.}

\label{colorazzo}
\end{figure}

\end{document}